\documentclass[aps,prd,preprint,groupedaddress]{revtex4-1}
\usepackage{graphicx}
\usepackage{amsmath}
\usepackage{bm}

\begin{document}
\begin{flushright}
  OU-HET-719 \ \\
\end{flushright}
\vspace{0mm}


\title{On the electromagnetic hadron current derived from \\
the gauged Wess-Zumino-Witten action}


\author{M.~Wakamatsu}
\email[]{wakamatu@phys.sci.osaka-u.ac.jp}
\affiliation{Department of Physics, Faculty of Science, \\
Osaka University, \\
Toyonaka, Osaka 560-0043, JAPAN}



\begin{abstract}
In a recent paper, based on the Skyrme model with the
Wess-Zumino-Witten term including
electromagnetism, Eto et al. pointed out an intriguing possibility that
the Gell-Mann-Nishijima relation is corrected under the presence of
back-ground electromagnetic fields, thereby being led to the
conclusion that even a neutron acquires non-zero net charge in
external magnetic fields. We point out that this remarkable conclusion
is inseparably connected with an unwelcome feature of the gauged
Wess-Zumino-Witten action, i.e. the non-conservation of source
current of Maxwell equation. 
\end{abstract}



\maketitle


\section{Introduction}

In a recent paper  \cite{EHIIM11}, Eto et al. investigated electromagnetic
properties of baryons under the influence of external electromagnetic field,
based on the Skyrme model \cite{Skyrme61} with Wess-Zumino-Witten term
including electromagnetism \cite{WZ71},\cite{Witten83}, thereby
concluding that a nucleon in external electromagnetic
fields has anomalous charge distribution due to the chiral anomaly.
Furthermore, the Gell-Mann-Nishijima relation, $Q = I_3 + N_B / 2$
($Q$ : electric charge, $I_3$ : the third component of isospin,
$N_B$ : baryon number), acquires an additional term due to the quantum
anomaly. As a consequence, non-zero net charge, which is generally
non-integer, is induced even for a neutron. This astounding conclusion
stems from the gauged Wess-Zumino-Witten action with two flavors,
given in the form \cite{KRS84},\cite{PR85} :
\begin{equation}
 S_{WZW} [U, A_\mu] \ = \ - \ e \,\int \,d^4 x \,A_\mu \,
 \left(\, \frac{N_c}{6} \,j^\mu_B \ + \ \frac{1}{2} \,j^\mu_{anm} \,\right),
 \label{WZW1} 
\end{equation}
where
\begin{eqnarray}
 j^\mu_B &=& - \,\frac{1}{24 \,\pi^2} \,\epsilon^{\mu \nu \alpha \beta} \,
 \mbox{\rm tr} \,(L_\nu \,L_\alpha \,L_\beta) , \label{current_B} \\
 j^\mu_{anm} &=& \frac{i \,e \,N_c}{96 \,\pi^2} \,
 \epsilon^{\mu \nu \alpha \beta} \,F_{\nu \alpha} \,
 \mbox{\rm tr} \,\tau_3 \,(L_\beta + R_\beta) , \label{current_anm}
\end{eqnarray}
with
\begin{equation}
 L_\mu \ \equiv \ U \,\partial_\mu U^\dagger, \ \ \  
 R_\mu \ \equiv \ \partial_\mu U^\dagger \,U . 
\end{equation}
(We point out that our definition of $L_\mu$ and $R_\mu$ is different
from that in \cite{EHIIM11}.)
Here, $j^\mu_B$ is the well-known baryon current giving an integer baryon
number \cite{Witten83}. According to \cite{EHIIM11}, in the presence of
background electromagnetic fields, not only
the first term but also the second term of Eq.~(\ref{WZW1}) is important.
The electric charge $Q$ with the contribution of anomaly is then written as
\begin{equation}
 Q \ = \ I_3 \ + \ \frac{N_B}{2} \ + \ \frac{Q_{anm}}{2} ,
\end{equation}
where $N_B = \int d^3 x \,j^0_B$ and $Q_{anm} = \int \,d^3 x \,j^0_{anm}$.
This means that the Gell-Mann-Nishijima relation receives
a remarkable modification under the background electromagnetic fields.

It appears to us, however, that the above-mentioned anomalous induction
of non-zero net charge for a nucleon (or a Skyrmion) is not in good harmony
with the schematic physical picture illustrated in Fig.1 of their paper.
This schematic diagram represents electric charge generation of a nucleon
through the anomalous coupling between one pion and two photons
(or electromagnetic fields). Since the electromagnetic fields (as abelian
gauge fields) carries no electric charge, the exchanged pion in this
figure must be neutral. In fact, this lowest order diagram results
from the same vertex as describing the famous decay process
$\pi^0 \rightarrow 2 \,\gamma$ due to the triangle
anomaly \cite{Adler69},\cite{BJ69},
which is legitimately contained in the
gauged Wess-Zumino-Witten action \cite{Witten83}\nocite{KRS84}-\cite{PR85}.
Naturally, the gauged Wess-Zumino-Witten
action also contains higher-power terms in the pion fields.
However, even if one considers diagrams in which more pions are exchanged
between the nucleon and the electromagnetic fields, the exchanged pions
must be electrically neutral as a whole, since the electromagnetic fields
carry no electric charge. 
What we are worrying about here is a conflict between this intuitive thought
and the principle conclusion of the paper \cite{EHIIM11}, i.e. the anomalous
induction of non-zero net charge for a nucleon.

The purpose of the present paper is to unravel the origin of this contradiction.
Here, we unavoidably encounter the problem of how to define electromagnetic
hadron current in an unambiguous manner by starting with the gauged
Wess-Zumino action.
A subtlety arises from the fact that the gauged Wess-Zumino Witten action
contains nonlinear terms in the electromagnetic fields. In fact, if it contains
only linear terms in the electromagnetic fields, it is clear that one can easily
read off the electromagnetic hadron current as a coefficient of the
electromagnetic field. For handling this delicate point, 
first in sect.II, we briefly analyze the familiar lagrangian of
scalar electrodynamics containing couplings between photons and
complex scalar fields, which is nonlinear in the photon fields.
A particular emphasis here is put on how to define electromagnetic matter
current based on a solid guiding principle. It will be shown there that the two
forms of current, i.e. the one defined on the basis of the Noether theorem
and the other defined as a source current of the Maxwell equation through
the equations of motion, perfectly coincides with each other.
It is also shown that this current is gauge-invariant and conserved,
thereby ensuring the consistency of scalar electrodynamics as a quantum
gauge theory. In section III, we shall carry out a similar analysis for the
nonlinear meson action with the Wess-Zumino-Witten action including
electromagnetism to find something unexpected, which is thought to be
the origin of the discrepancy pointed out above. Finally, in sect. IV,
we briefly summarize what we have found in the present paper.

\section{A lesson learned from scalar electrodynamics}

Let us start with the familiar lagrangian of scalar electrodynamics given by
\begin{equation}
 {\cal L} \ = \ - \,\frac{1}{4} \,F_{\mu \nu} \,F^{\mu \nu} \ + \ 
 D_\mu \phi^* \,D^\mu \phi \ - \ V (\phi^* \,\phi) ,
\end{equation}
with
\begin{eqnarray}
 F_{\mu \nu} &\equiv& \partial_\mu \,A_\nu \ - \ 
 \partial_\nu A_\mu , \\
 D^\mu \,\phi (x) &\equiv& [\,\partial^\mu \ + \ i \,e \,A^\mu (x) \,] \,
 \phi(x) .
\end{eqnarray}
This lagrangian is manifestly gauge-invariant under the following gauge
transformation : 
\begin{eqnarray}
 \phi (x) &\rightarrow& e^{- \,i \,e \,\alpha (x)} \,\phi (x), \ \ \ 
 A^\mu (x) \ \rightarrow \ A^\mu (x) \ + \ \partial^\mu \,\alpha (x).
 \label{SED_GT}
\end{eqnarray}
The equations of motion derived from the above lagrangian are given by
\begin{eqnarray}
 \partial_\mu \,F^{\mu \nu} &=& j^\nu , \label{SED_Maxwell_eq}\\
 D_\mu \,D^\mu \,\phi &=& - \,\partial V \,/ \,\partial \phi^*, \\
 \left( D_\mu \,D^\mu \,\phi \right)^* &=& - \,\partial V \,/ \,\partial \phi .
\end{eqnarray}
Here, the source current $j^\nu$ of the Maxwell
equation (\ref{SED_Maxwell_eq}) is given by
\begin{equation}
 j^\nu \ = \ i \,e \,\left[\, \phi^* \,D^\mu \,\phi \ - \ 
 (D^\mu \,\phi)^* \,\phi \,\right] \ = \ 
 i \,e \,\phi^* \,\overleftrightarrow{\partial^\nu} \,\phi \ - \ 
 2 \,e^2 \,\phi^* \,\phi \,A^\nu , 
\end{equation}
with $\overleftrightarrow{\partial^\nu} = \overrightarrow{\partial^\nu} - \overleftarrow{\partial^\nu}$.
By using the equations of motion, it can be shown that this matter current
$j^\nu$ is conserved, i.e.
\begin{equation}
 \partial_\nu \,j^\nu \ = \ 0 .
\end{equation}
One can also convince that this matter (or source) current is
invariant under the full gauge transformation (\ref{SED_GT}).
One should recognize that the conservation of source current is crucially
important. In fact, if it were broken, one encounters a serious contradiction
with the Maxwell equation (\ref{SED_Maxwell_eq}) in such a way that
\begin{equation}
 0 \ = \ \partial_\nu \,\partial_\mu \,F^{\mu \nu} \ = \ 
 \partial_\nu \,j^\nu \ \neq \ 0.
\end{equation}
For later discussion, it is useful to remember the fact that the matter current
above can also be obtained by using the standard Noether prescription.
To confirm it, we first consider the infinitesimal version of the gauge
transformation given by
\begin{eqnarray}
 \delta \phi \ = \ - \,i \,e \,\epsilon (x) \, \phi , \ \ \ 
 \delta \phi^* \ = \ i \,e \,\epsilon (x) \,\phi^*, \ \ \ 
 \delta A^\mu \ = \ \partial^\mu \epsilon (x).
\end{eqnarray}
Naturally, the lagrangian of the scalar electrodynamics is invariant under
this gauge transformation.
The Noether current is obtained by considering another variation
\begin{eqnarray}
 \delta^\prime \phi \ = \ - \,i \,e \,\epsilon (x) \, \phi , \ \ \ 
 \delta^\prime \phi^* \ = \ i \,e \,\epsilon (x) \,\phi^*, \ \ \ 
 \delta^\prime  A^\mu \ = \ 0.
\end{eqnarray}
The variation of the whole lagrangian under this transformation
is reduced to the form
\begin{equation}
 \delta^\prime {\cal L} \ = \ \epsilon (x) \,\partial_\mu \,J^\mu \ + \ 
 \partial_\mu \epsilon (x) \,J^\mu,
\end{equation} 
which defines the current $J^\mu$ such a way that 
\begin{eqnarray}
 J^\mu \ &=& \frac{\partial (\delta^\prime {\cal L})}{\partial (\partial_\mu \epsilon (x))}, 
 \label{SED_Noether_current} \\
 \partial_\mu \,J^\mu &=& \ \frac{\partial (\delta^\prime {\cal L})}{\partial \epsilon (x)}.
\end{eqnarray} 
If the lagrangian ${\cal L}$ is invariant under a space-time independent
(global) transformation
\begin{eqnarray}
 \delta^\prime \phi \ = \ - \,i \,e \,\epsilon \, \phi , \ \ \ 
 \delta^\prime \phi^* \ = \ i \,e \,\epsilon \,\phi^*, 
\end{eqnarray}
which is indeed the case with our lagrangian (1), we conclude that
\begin{equation}
 0 \ = \ \delta^\prime {\cal L} \ = \ \epsilon \,\,\partial_\mu \,J^\mu.
\end{equation}
This means that the current $J^\mu$ defined by the equation
(\ref{SED_Noether_current}) is in fact
a conserved Noether current. The above-explained method of obtaining
the Noether current is known as the Gell-Mann-Levy method.

Now, for the lagrangian of the scalar electrodynamics, we have
\begin{eqnarray}
 \delta^\prime [\,\partial_\mu \,\phi^* \,\partial^\mu \phi \,]
 &=& i \,e \,\partial_\mu \,\epsilon (x) \,
 [\,\phi^* \,\partial^\mu \,\phi \ - \ \partial^\mu \,\phi^* \,\phi \,], \\
 \delta^\prime \,[\,i \,e \,(\,\partial_\mu \,\phi^* \,\phi \ - \ 
 \phi^* \,\partial_\mu \,\phi \,) \,A^\mu \,] &=&
 \partial_\mu \epsilon (x) \,( - \,2 \,e^2 ) \,\phi^* \,\phi \,A^\mu , \\
 \delta^\prime \,[\,e^2 \,A_\mu \,A^\mu \,\phi^* \,\phi \,] &=& 0 ,
\end{eqnarray}
thereby being led to
\begin{equation}
 \delta^\prime \,{\cal L} \ = \ \partial_\mu \,\epsilon (x) \,
 \left\{\,i \,e \,[\,\phi^* \,\partial^\mu \,\phi \ - \ 
 (\partial^\mu \,\phi)^* \,\phi \,]
 \ - \ 2 \,e^2 \,\phi^* \,\phi \,A^\mu \,\right\} .
\end{equation}
The resultant Noether current is therefore given by
\begin{eqnarray}
 J^\mu &=& i \,e \,\,\phi^* \,\overleftrightarrow{\partial^\mu} \,\phi 
 \ - \ 2 \,e^2 \,\phi^* \,\phi \,A^\mu
 \ = \ 
 i \,e \,[\,\phi^* \,D^\mu \,\phi \ - \ (D^\mu \,\phi)^* \,\phi \,].
\end{eqnarray}
One confirms that this conserved Noether current precisely coincides
with the source current appearing in the Maxwell
equation (\ref{SED_Maxwell_eq}) for the photon
field. This ensures the consistency of the scalar electrodynamics as a
classical and a quantum field theory. Somewhat embarrassingly, we shall
see below that the familiar gauged Wess-Zumino-Witten action does not
satisfy the same sense of consistency.

\section{electromagnetic hadron current resulting from the gauged
Wess-Zumino-Witten action}

\subsection{Matter current derived from Noether principle}

Here, we start with the gauged Wess-Zumino-Witten action with two flavor
expressed in the following form :
\begin{eqnarray}
 S_{WZW} [\,U, A_\mu \,] &=& S_{WZ} [U] \ - \ 
 \frac{1}{2} \,e \,\int \,d^4 x \,A_\mu
 \nonumber \\
 &\times& \left\{\,- \,\frac{1}{24 \,\pi^2} \,
 \epsilon^{\mu \nu \alpha \,\beta} \,
 \mbox{tr} \,(L_\nu \, L_\alpha \,L_\beta ) \right. \nonumber \\
 &\,& \ \ \left. + \ \frac{3 \,i \,e}{48 \,\pi^2} \,
 \epsilon^{\mu \nu \alpha \beta} \,
 F_{\nu \alpha} \,\mbox{tr} \,Q \,(L_\beta + R_\beta) \,\right\} ,
 \label{WZW2}
\end{eqnarray}
where $Q$ is the SU(2) charge matrix given as
\begin{equation}
 Q \ = \ \left( \begin{array}{cc}
 \frac{2}{3} & 0 \\
 0 & - \,\frac{1}{3} \\
 \end{array} \right) .
\end{equation}
(As is well-known, although $S_{WZ} [U]$ vanishes in the SU(2) case,
its gauge variation does not. We therefore retain it here.) 
By construction, i.e. as a consequence of the
``trial and error'' gauging {\it a la} Witten \cite{Witten83}, the gauged
Wess-Zumino-Witten action is invariant under the following
infinitesimal gauge transformation : 
\begin{equation}
 \delta U \ = \ i \,\epsilon (x) \,[Q, U] , \ \ 
 \delta U^\dagger \ = \ i \,\epsilon (x) \,[Q, U^\dagger], \ \ 
 \delta \,A_\mu \ = \ - \,\frac{1}{e} \,\,\partial_\mu \,\epsilon (x) .
 \label{fullGT}
\end{equation}
Let us first try to see what answer we shall obtain for the Noether current,
if we apply the Gell-Mann-Levy method to the above lagrangian (\ref{WZW2}).
The transformation, which we consider to this end, is given by
\begin{equation}
 \delta^\prime U \ = \ i \,\epsilon (x) \,[Q, U] , \ \ 
 \delta^\prime U^\dagger \ = \ i \,\epsilon (x) \,[Q, U^\dagger], \ \ 
 \delta^\prime \,A_\mu \ = \ 0 . 
\end{equation}
Making use of the relation
\begin{eqnarray}
 \delta^\prime \,S_{WZW} &=& - \,\int \,d^4 x \,\,\partial_\mu \epsilon (x) \,
 \frac{1}{48 \,\pi^2} \,\epsilon^{\mu \nu \alpha \beta} \,\mbox{tr} \,
 (L_\nu \,L_\alpha \,L_\beta ) \nonumber \\
 &\,& - \ \int \,d^4 x \,\,\partial_\nu \,\epsilon (x) \,\frac{3 \,i \,e}{48 \,\pi^2}
 \,\epsilon^{\mu \nu \alpha \beta} \,A_\mu \,\partial_\alpha \,
 \mbox{tr} \,Q \,(L_\beta + R_\beta),
\end{eqnarray}
we readily find that the corresponding Noether current is given by
\begin{eqnarray}
 J^\mu_{\rm I} \ \equiv \ \frac{\delta \,(\delta^\prime \,S_{WZW})}
 {\delta (\partial_\mu \,\epsilon (x))} &=&
 - \,\frac{1}{48 \,\pi^2} \,\,\epsilon^{\mu \nu \alpha \beta} \,\,
 \mbox{tr} \,(L_\nu \,L_\alpha \,L_\beta) \nonumber \\
 &\,& + \,\frac{3 \,i \,e}{48 \,\pi^2} \,\epsilon^{\mu \nu \alpha \beta} \,\,
 A_\nu \,\partial_\alpha \,
 \mbox{tr} \,Q \,(L_\beta + R_\beta) .
\end{eqnarray}
One can also verify that this current is invariant under the full
gauge transformation (\ref{fullGT}). Unfortunately, this current is not conserved.
In fact, we find that
\begin{eqnarray}
 \partial_\mu \,J^\mu_{\rm I} &=& \frac{3 \,i \,e}{24 \,\pi^2} \,\,
 \epsilon^{\mu \nu \alpha \beta} \,\partial_\mu A_\nu \,\,
 \partial_\alpha \,\mbox{tr} \,Q \,(L_\beta + R_\beta) \ \neq \ 0. \label{div_current}
\end{eqnarray}
However, one can verify that the r.h.s. of (\ref{div_current}) is a total
derivative of another four-vector as
\begin{equation}
 \partial_\mu \,J^\mu_{\rm I} \ = \ \partial_\mu \,X^\mu,
\end{equation}
with
\begin{equation}
 X^\mu \ \equiv \ \frac{3 \,i \,e}{48 \,\pi^2} \,
 \epsilon^{\mu \nu \alpha \beta} \,A_\nu \,\partial_\alpha \,
 \mbox{tr} \,Q \,(L_\beta + R_\beta) .
\end{equation}
This means that, if we define another current $j^\mu_{\rm II}$ by
\begin{equation}
 J^\mu_{\rm II} \ \equiv \ J^\mu_{\rm I} \ - \ X^\mu \ = \ 
 - \,\frac{1}{24 \,\pi^2} \,\epsilon^{\mu \nu \alpha \beta} \,
 \mbox{tr} \,(L_\nu \,L_\alpha \,L_\beta ),
\end{equation}
then, $J^\mu_{\rm II}$ is conserved. The price to pay is that the new current
$J^\mu_{\rm II}$ is no longer gauge-invariant.

Incidentally, in the case of Poincare symmetry not of internal symmetry,
the ambiguous nature of the Noether current is widely known.
For example, in the case of quantum
chromodynamics (QCD), the 2nd-rank energy momentum tensor obtained from
a naive Noether procedure does not satisfy the desired symmetry property
under the exchange of two Lorentz indices \cite{JM90}. However, there exists
a well-known procedure for ``improving'' the Noether current by adding
a superpotential - divergence of anti-symmetric tensor - which does not
spoil the current conservation. The symmetric energy momentum tensor
of QCD obtained in such a procedure is sometimes called Belinfante
symmetrized energy-momentum tensor. 

Summarizing the analysis in this subsection, we have applied the familiar
Gell-Mann-Levy method to the gauged Wess-Zumino-Witten action for
obtaining a Noether current as a candidate of electromagnetic hadron
current. However, we have ended up
with two different forms of currents, i.e. $J^\mu_{\rm I}$ and $J^\mu_{\rm II}$.
The current $J^\mu_{\rm I}$ is gauge-invariant but not conserved, while
the current $J^\mu_{\rm II}$ is conserved but not gauge-invariant.
As pointed out in the paper by Son and Stephanov \cite{SS08}, one can
construct the 3rd current, which satisfies both of gauge-invariance and current
conservation, by using the ``trial and error'' gauging method as proposed by
Witten. It is given by
\begin{eqnarray}
 J^\mu_{\rm III} &=& - \,\frac{1}{48 \,\pi^2} \,\epsilon^{\mu \nu \alpha \beta} \,
 \mbox{tr} \,(L_\nu \,L_\alpha \,L_\beta) \nonumber \\
 &\,& - \,\frac{3 \,i \,e}{48 \,\pi^2} \,\epsilon^{\mu \nu \alpha \beta} \,
 \partial_\nu \,[\,A_\alpha \,\mbox{tr} \,Q \,(L_\beta + R_\beta) \,].
\end{eqnarray}
Unfortunately, it is not a current derived from the gauged Wess-Zumino-Witten
action on the basis of a definite prescription as guidelined by the Noether principle.

In this way, we must conclude that, quite different from the case of
scalar electrodynamics, the standard Noether method does not do a
desired good job to derive the electromagnetic hadron current
corresponding to the gauged Wess-Zumino-Witten action, in the sense 
that it fails to give a candidate of electromagnetic hadron current,
satisfying both of gauge-invariance and conservation.
In the next subsection, we shall investigate the nature of another candidate
of electromagnetic hadron current, i.e. the source current, which is defined
through the equation of motion for the electromagnetic field.

\subsection{Matter current as a source of Maxwell equation}

The full action of the two-flavor Skyrme model coupled to the
electromagnetic field $A_\mu$ is given by
\begin{equation}
 S \ = \ S_\gamma [A_\mu] \ + \ S_{Skyrme} [U, A_\mu] \ + \ 
 S_{WZW} [U, A_\mu] .
\end{equation}
Here, the 1st term
\begin{equation}
 S_\gamma \ = \ - \,\frac{1}{4} \,\int \,d^4 x \,\,F_{\mu \nu} \,F^{\mu \nu}
\end{equation}
is the kinetic part of the electromagnetic field, while the 2nd term,
$S_{Skyrme} [U,A_\mu]$, stands for the non-anomalous part of action
for the two-flavor Skyrme model minimally coupled to the electromagnetic
field. The 3rd term, i.e. $S_{WZW} [U, A_\mu]$, gives the gauged
Wess-Zumino-Witten action given by (\ref{WZW2}).
In the following, we shall discard
the part $S_{Skyrme} [U, A_\mu]$ for simplicity, since it plays no essential role
in our discussion below. The Euler-Lagrange equation of motion for the
electromagnetic field is therefore written down from
\begin{equation}
 \frac{\delta}{\delta A_\nu} \,\left\{\,S_\gamma [A_\mu] \ + \ 
 S_{WZW} [U, A_\mu] \,\right\} \ = \ 0 ,
\end{equation}
which gives the Maxwell equation
\begin{equation}
 \partial_\mu \,F^{\mu \nu} \ = \ j^\nu ,
\end{equation}
with the definition of the source current $j^\nu$ as
\begin{equation}
 e \,j^\nu \ \equiv \ - \,\frac{\delta}{\delta A_\nu} \,S_{WZW} [U, A_\mu] .
\end{equation}
(Naturally, if we had included the part $S_{Skyrme} [U, A_\mu]$, it would also
contribute to the source current of Maxwell equation. However, this part
of current is conserved itself and does not cause any trouble as discussed
below.) An immediate question is whether the above definition, given as a
functional derivative of the gauged Wess-Zumino-Witten action with respect
to the electromagnetic fields, offers us the same answer as obtained with
the Noether prescription. The answer is no.
We find that the source current is given by
\begin{eqnarray}
 j^\mu &=& - \,\frac{1}{48 \,\pi^2} \,\epsilon^{\mu \nu \alpha \beta} \,\,
 \mbox{\rm tr} \,(L_\nu \,L_\alpha \,L_\beta) \nonumber \\
 &\,& + \,\frac{3 \,i \,e}{96 \,\pi^2} \,\epsilon^{\mu \nu \alpha \beta} \,\,
 F_{\nu \alpha} \,\mbox{\rm tr} \,Q \,(L_\beta + R_\beta) \nonumber \\
 &\,& + \,\frac{3 \,i \,e}{48 \,\pi^2} \,\,\epsilon^{\mu \nu \alpha \beta} \,\,
 \partial_\nu \,\left[\,
 A_\alpha \,\mbox{\rm tr} \,Q \,(L_\beta + R_\beta) \,\right] . 
 \label{source_current}
\end{eqnarray}
which does not coincide with any of the currents
$J^\mu_{\rm I}$, $J^\mu_{\rm II}$, and $J^\mu_{\rm III}$ discussed in the
previous subsection. Somewhat unexpectedly, it turns out that this current
$j^\mu$ is not gauge-invariant. More serious problem is that it
is not conserved, owing to the presence of the 2nd term
of (\ref{source_current}). In fact, we find that
\begin{equation}
 \partial_\mu \,j^\mu \ = \ \partial_\mu \,X^\mu \ \neq \ 0,
\end{equation}
with
\begin{equation}
 X^\mu \ = \ \frac{3 \,i \,e}{48 \,\pi^2} \,\epsilon^{\mu \nu \alpha \beta} \,\,
 A_\nu \,\partial_\alpha \,\mbox{\rm tr} \,Q \,(L_\beta + R_\beta) .
\end{equation}
As emphasized in the example of scalar electrodynamics, non-conservation
of source current is not permissible, since it causes an incompatibility
with the fundamental equation of electromagnetism, i.e.
the Maxwell equaion \cite{Jackiw85}.
How can we make a compromise with this trouble. One possible attitude would
be to follow the argument as given by Kaymakcalan, Rajeev and Schechter many
years ago \cite{KRS84}.
They argue that the low-energy effective action for QCD involves many more
new fields and interactions so one should not worry too much about the
complete consistency of equation of motion. The effective action is, after all,
being used as a handy mnemonic to read off the relevant vertices. The
gauged Wess-Zumino-Witten action certainly describes the typical
anomalous processes containing the photons like $\pi^0 \rightarrow 2 \,\gamma$
and/or $\gamma \rightarrow 3 \,\pi$ consistently with the low energy theorem,
i.e. the anomalous Ward identities.

Now we are in a position to pinpoint the origin of somewhat astounding conclusion
obtained in the paper \cite{EHIIM11}, i.e. the anomalous induction of net electric
charge for a nucleon in the magnetic fields. This conclusion follows from
the electromagnetic hadron current given as a half of the sum
of $j^\mu_B$ in (\ref{current_B}) and $j^\mu_{anm}$ in (\ref{current_anm}).
Setting $N_c = 3$, this reduces to
\begin{eqnarray}
 j^\mu &=& - \,\frac{1}{48 \,\pi^2} \,\epsilon^{\mu \nu \alpha \beta} \,\,
 \mbox{\rm tr} \,(L_\nu \,L_\alpha \,L_\beta) \nonumber \\
 &\,& + \,\frac{i \,e \,N_c}{192 \,\pi^2} \,\epsilon^{\mu \nu \alpha \beta} \,\,
 F_{\nu \alpha} \,\,\mbox{\rm tr} \,\,\tau_3 \,(L_\beta + R_\beta) .
 \label{current_B+anm}
\end{eqnarray}
In consideration of the fact that $Q = \frac{1}{6} + \frac{\tau_3}{2}$,
this current just coincides with the sum of the 1st and 2nd terms
in the source current (47), which we have derived above.
Since the 3rd term of the current (\ref{source_current}) is of a
total derivative form, it does not contribute to the net charge of a
nucleon. We thus find that the 2nd term of the current (\ref{source_current})
or of the current (\ref{current_B+anm}) is
the cause of trouble, which prevents
the conservation of source current of the Maxwell equation.
In any case, what we can say definitely from the analysis above is that
the anomalous induction of non-zero net charge for a nucleon (or a Skyrmion)
claimed in the paper \cite{EHIIM11} is inseparably connected with
this unfavorable feature of the gauged Wess-Zumino-Witten action.
Still, what is lacking in our understanding is a deep explanation of why the
gauged Wess-Zumino-Witten action, which was constructed so as to fulfill the
electromagnetic gauge-invariance with use of the ``trial and error'' method,
does not satisfy the consistency with the Maxwell equation. 
 
A final comment is on a related work by Kharzeev, Yee, and Zahed \cite{KEZ11},
which was done motivated by the paper \cite{EHIIM11}.
Starting with a simple effective lagrangian of QCD (it corresponds to
the lowest power term in the pion field in the gauged Wess-Zumino-Witten
action), they investigated the effect of anomaly induced charge distribution
in the nucleon. Under a certain kinematical approximation concerning the
classical equation of motion for the pion field in a nucleon, they
conclude that the abelian anomaly of QCD induces a quadrupole moment
for a neutron but it does not induce net electric charge for it.
The last statement, i.e. no induction of net electric charge for a neutron
appears to be consistent with the nature of their effective lagrangian
and also with the intuitive consideration given in the introduction of
the present paper.

\section{Summary and conclusion}

To conclude, motivated by the recent claim that, under the external magnetic
fields, the anomalous couplings between mesons and electromagnetic fields
contained in the gauged Wess-Zumino-Witten action induces non-zero
net electric-charge for a nucleon, we have carefully re-investigated the
problem of how to define the electromagnetic hadron current from this
widely-known action. To this end, we first compare the two methods of
obtaining electromagnetic hadron current for the familiar lagrangian
of the scalar electrodynamics. The one is the Gell-Mann-Levy
method to obtain the Noether current, while the other is the method of
using equations of motion of actions to define source current.
For this standardly-known lagrangian, we confirm that these
two methods give precisely the same form of the electromagnetic
hadron current. It can also be verified that this current is gauge-invariant
and conserved. Unfortunately, this is not the case with
the gauged Wess-Zumino-Witten action.
That is, the currents obtained by these two methods do not coincide with
each other. Particularly troublesome here is the fact that the source current
of Maxwell equation is not conserved. This means that
the gauged Wess-Zumino-Witten action, which was constructed so as to
fulfill the electromagnetic gauge-invariance by using the ``trial and error''
method, does not satisfy the consistency with the the fundamental equation
of electromagnetism. Although mysterious, it seems at the least
clear that the recently claimed anomalous induction of net electric charge
for a nucleon in the magnetic fields is inseparably connected with this
unwelcome feature of the gauged Wess-Zumino-Witten action. 

\begin{acknowledgments}
The author would like to thank Prof. T.~Kubota for useful discussion.
This work is supported in part by a Grant-in-Aid for Scientific
Research for Ministry of Education, Culture, Sports, Science
and Technology, Japan (No.~C-21540268)
\end{acknowledgments}


\end{document}